\documentclass[aps,prl,twocolumn,superscriptaddress,showpacs,amsmath,amssymb]{revtex4}

\usepackage{graphicx}
\usepackage{dcolumn}
\usepackage{bm}

\begin{document}


\title{
Spin Fluctuations and 
Non-Fermi-Liquid Behavior of 
CeNi$_2$Ge$_2$
}

\author{H. Kadowaki}  
\affiliation{Department of Physics, Tokyo Metropolitan University, 
Hachioji-shi, Tokyo 192-0397, Japan}

\author{B. F{\aa}k}
\affiliation{CEA-Grenoble, DSM/DRFMC/SPSMS, 38054 Grenoble, France}
\affiliation{ISIS Facility, Rutherford Appleton Laboratory,
Chilton, Didcot, OX11 0QX, United Kingdom}

\author{T. Fukuhara}  
\affiliation{Department of Liberal Arts and Science, 
Toyama Prefectural University, 
Kosugi, Toyama 939-0398, Japan}

\author{K. Maezawa}  
\affiliation{Department of Liberal Arts and Science, 
Toyama Prefectural University, 
Kosugi, Toyama 939-0398, Japan}

\author{K. Nakajima}   
\affiliation{ISSP-NSL, University of Tokyo, 
106-1 Shirakata, Tokai, Ibaraki 319-1106, Japan}

\author{M. A. Adams}
\affiliation{ISIS Facility, Rutherford Appleton Laboratory,
Chilton, Didcot, OX11 0QX, United Kingdom}

\author{S. Raymond}
\affiliation{CEA-Grenoble, DSM/DRFMC/SPSMS, 38054 Grenoble, France}

\author{J. Flouquet}
\affiliation{CEA-Grenoble, DSM/DRFMC/SPSMS, 38054 Grenoble, France}

\date{\today}

\begin{abstract}
Neutron scattering shows that non-Fermi-liquid behavior of the heavy-fermion compound 
CeNi$_2$Ge$_2$ 
is brought about by the development of low-energy spin fluctuations with an energy scale of 0.6 meV. 
They appear around the antiferromagnetic wave vectors 
$(\frac{1}{2} \frac{1}{2} 0)$ and $(0 0 \frac{3}{4})$ 
at low temperatures, 
and coexist with high-energy spin fluctuations with an energy scale of 4 meV and a modulation vector 
$(0.23, 0.23, \frac{1}{2})$. 
This unusual energy dependent structure of 
$\text{Im} \chi (\bm{Q}, E)$ 
in $\bm{Q}$ 
space suggests that quasiparticle bands are important.
%
\end{abstract}

\pacs{
75.30.Mb, 71.10.Hf, 71.27.+a
}


\maketitle

Non-Fermi-liquid (NFL) behavior has been investigated in an increasing 
number of $d$- and $f$-electron systems in recent years 
\cite{Stewart01,Coleman01}. 
In usual heavy-fermion systems, although strong correlation effects of 
$f$ electrons bring about a mass renormalization $m^*/m$ by a factor of up 
to a few thousands, the systems remain in Fermi liquid (FL) states, 
which are typically observed as $C/T$ = const and 
$\rho - \rho_0 \propto T^2$ at low temperatures. 
The large mass enhancement originates from fluctuations of the spin degrees of 
freedom of the $f$ electrons participating in the quasiparticles. 
When spin fluctuations are slowed down by certain mechanisms, the FL 
description breaks down, and NFL behavior appears as, for example, 
$C/T \propto \ln (T_0 / T)$ and 
$\rho - \rho_0 \propto T^x$ with $x<2$.

A mechanism of NFL behavior is critical spin fluctuations near a quantum 
critical point (QCP), i.e., a zero-temperature magnetic phase transition, 
$T_{\text{N}}$ (or $T_{\text{C}}$) = 0
\cite{Moriya-Taki,Hertz76-Millis93,Coleman01}. 
Observation of a QCP requires tuning of the competition between quenching 
of spin by the Kondo effect and interspin coupling by 
Ruderman-Kittel-Kasuya-Yosida (RKKY) interactions using chemical 
substitutions, static pressures, or magnetic fields \cite{Lohneysen94}. 
Recent experimental studies on critical behavior of CeCu$_{5.9}$Au$_{0.1}$ 
\cite{Lohneysen94,Schroder00} 
posed an intriguing theoretical question: Is the singularity 
described by the standard spin-fluctuation theories 
\cite{Moriya-Taki,Hertz76-Millis93}
or a locally critical quantum phase transition \cite{Coleman01,Si01}?
For chemically substituted systems, disorders inevitably affect 
singularities, ranging from perturbative effects to disorder-driven NFL 
behaviors \cite{disorder-NFL}. 
Experiments using stoichiometric compounds showing NFL behavior 
without tuning, such as CeNi$_2$Ge$_2$ \cite{Steglich96-Julian96} 
and YbRh$_2$Si$_2$ \cite{Trovarelli00}, are thus expected to clarify the QCP 
or other mechanisms of NFL in the clean limit. 

\begin{figure}
\begin{center}
\includegraphics[width=8.7cm,clip]{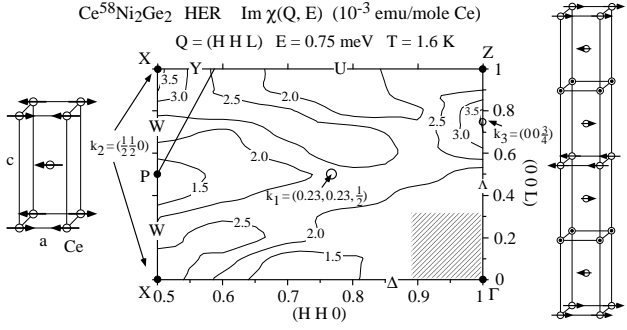}
\end{center}
\caption{
\label{fig:map}
A contour map of constant-$E$ scans taken with $E = 0.75$ meV in the $(HHL)$ 
scattering plane at $T = 1.6$ K. 
No data were taken in the hatched area,  due to nonmagnetic background. 
Possible antiferromagnetic spin configurations, depicted on the left and right 
sides, illustrate low-energy spin fluctuations with wave vectors 
$\bm{k}_2 = (\frac{1}{2} \frac{1}{2} 0)$ ($X$ point) and 
$\bm{k}_3 = (0 0 \frac{3}{4})$, respectively, 
assuming spins along the $a$ axis. 
The wave vector $\bm{k}_1 = (0.23, 0.23, \frac{1}{2})$ is 
the position where the high-energy spin fluctuation 
($\Gamma_{\bm{Q}} \simeq 4$ meV) shows 
maximum intensity \cite{Fak-JPCM}.
}
\end{figure}

CeNi$_2$Ge$_2$, which crystallizes in a body-centered 
tetragonal structure (see Fig.~\ref{fig:map}), is a paramagnetic 
heavy-fermion compound with enhanced $C/T \simeq 350$ mJ/K$^2$ mol 
\cite{Knopp88}. 
It shows Kondo behavior with a temperature scale of 
$T_{\text{K}} \simeq 30$ K \cite{Knopp88}
and has a metamagnetic behavior at 
$H_{\text{M}} \simeq 42$ T \cite{Fuku-MetaM}. 
For $T < 5$ K, i.e.,  well below $T_{\text{K}}$, CeNi$_2$Ge$_2$ exhibits
NFL behavior with $C/T \propto \ln (T_0 / T)$ and $\rho - \rho_0 \propto T^x$, 
where $1<x<1.5$ \cite{Steglich96-Julian96}. 
CeNi$_2$Ge$_2$ also displays superconductivity near the 
QCP \cite{Grosche00-Braithwaite00} which may be 
spin-fluctuation mediated \cite{Mathur98}. 

The NFL behavior has been thought to be caused by 
the spin fluctuations being 
slowed down by a QCP of an antiferromagnetic phase, 
which would be one of those observed in 
Pd, Rh, or Cu substituted compounds \cite{Sparn97,Fukuhara98,Knebel99}. 
However, previous neutron-scattering experiments \cite{Fak-JPCM}
on single crystalline CeNi$_2$Ge$_2$ 
disagree with this simple interpretation. 
The dynamical susceptibility is well described by the standard form 
\begin{equation}
\label{eq:L}
\text{Im} \chi_{\text{L}} (\bm{Q}, E) = 
\chi(\bm{Q}) \Gamma_{\bm{Q}} 
\frac{ E }{ E^2 + \Gamma_{\bm{Q}}^2 }
\; ,
\end{equation}
used in the spin-fluctuation theory \cite{Moriya-Taki}. 
However, the energy scale $\Gamma_{\bm{Q}} \sim 4$ meV 
$\sim \text{k}_{\text{B}} T_{\text{K}}$ 
shows only a weak $\bm{Q}$ dependence.
This is in contradiction with the QCP scenario, 
in which $\Gamma_{\bm{Q}}$ is expected to depend strongly on $\bm{Q}$ and 
vanish at the antiferromagnetic wave vector $\bm{k}_1 = (0.23, 0.23, \frac{1}{2})$
at $T = 0$. 

In this Letter, we present neutron-scattering measurements 
that reveal a second type of spin fluctuations, 
which are shown to be characterized by a lower-energy scale 
and highly relevant to the NFL behavior. 
The main part of the measurements was performed on the triple-axis 
spectrometer HER at JAERI,
equipped with a PG(002) monochromator 
and a horizontally focusing PG(002) analyzer. 
The typical energy resolution using a final energy of
$E_{f}=3.1$ meV was 0.1 meV 
[full width at half maximum (FWHM)] at the elastic position. 
Complementary measurements at lower energies were done on 
the IRIS time-of-flight spectrometer at RAL, 
with an energy resolution of 15 $\mu$eV (FWHM). 
Single crystals were grown by the Czochralski method 
using isotopic ${}^{58}$Ni, which is important to avoid the 
large incoherent elastic scattering of natural Ni. 
Four crystals with a total volume of 2.2 cm$^3$ were 
aligned together and 
mounted in He flow  or dilution cryostats. 
All the data shown are converted to the dynamical susceptibility 
and corrected for the magnetic form factor. 
It is scaled to absolute units by comparison with the intensity of 
the incoherent scattering from a vanadium sample. 

A number of constant-$E$ scans covering 
an irreducible Brillouin zone were performed to search for 
low-energy spin fluctuations at $T = 1.6$ K. 
Constant-$E$ scans at $E = 0.75$ meV 
in the $(HHL)$ scattering plane show (see Fig.~\ref{fig:map})
that there are two peak 
structures around 
$\bm{Q} = (\frac{1}{2} \frac{1}{2} 1)$ and 
$(1 1 \frac{3}{4})$, i.e., at
reduced wave vectors of 
$\bm{k}_2 = (\frac{1}{2} \frac{1}{2} 0)$ and 
$\bm{k}_3 = (0 0 \frac{3}{4})$. 
The wave vector $\bm{k}_2$ is the $X$ point in the Brillouin zone, 
which also corresponds to $\bm{Q} = (\frac{1}{2} \frac{1}{2} 0)$ 
in Fig.~\ref{fig:map}, where a smaller peak is seen. 
We note that strong intensities were observed only around 
$\bm{k}_2$ and $\bm{k}_3$ in the whole Brillouin zone, except for 
the vicinities of the 
$\Gamma$ point, where the high background prohibited us 
from measuring the magnetic scattering. 
Possible antiferromagnetic spin configurations modulated by $\bm{k}_2$ 
and $\bm{k}_3$ are illustrated on the left and right sides 
of the contour map, respectively, assuming that the spins are parallel to 
the $a$ axis. 

The spin-fluctuation scattering at $E = 0.75$ meV is peaked 
at the wave vectors $\bm{k}_2$ and $\bm{k}_3$, 
in contrast to that at $\sim 4$ meV, which is centered at 
$\bm{k}_1$ (see Fig.~\ref{fig:map}) and elongated in the [110] 
direction \cite{Fak-JPCM}. 
This feature cannot be accounted for by the spin-fluctuation 
theory of Ref.~\cite{Moriya-Taki}, 
since the product $\chi(\bm{Q}) \Gamma_{\bm{Q}}$ of
Eq.~(\ref{eq:L}) is predicted to be $\bm{Q}$ independent. 
A constant $\chi(\bm{Q}) \Gamma_{\bm{Q}}$ implies that 
$\text{Im} \chi_{\text{L}} (\bm{Q}, E)$ 
peaks at a $\bm{Q}$ vector where $\Gamma_{\bm{Q}}$ is minimum, 
which excludes the possibility to have other peaks in 
$\text{Im} \chi_{\text{L}} (\bm{Q}, E)$ at different energies. 
The archetypal heavy fermions 
CeRu$_2$Si$_2$ and CeCu$_6$, on the other hand, 
are in agreement with the 
$\bm{Q}$ independent product $\chi(\bm{Q}) \Gamma_{\bm{Q}}$ \cite{JRM88}.
The failure of the description of CeNi$_2$Ge$_2$ by the 
spin-fluctuation theory will be a clue to clarify its NFL behavior.

\begin{figure}
\begin{center}
\includegraphics[width=7.0cm,clip]{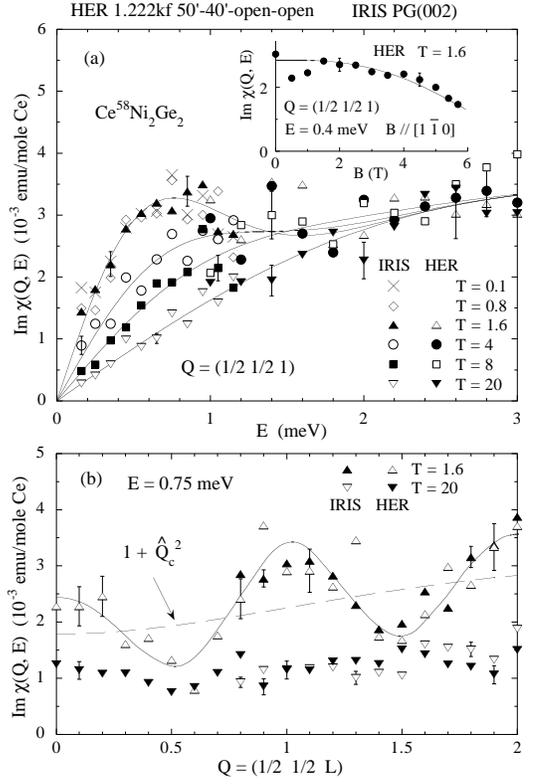}
\end{center}
\caption{
\label{fig:EQscan}
(a) Constant-$Q$ scans at 
$\bm{Q} = (\frac{1}{2} \frac{1}{2} 1)$. 
Solid lines are fits to the Lorentzian of Eq.~(\ref{eq:L}) 
with $\Gamma_{\bm{Q}} = 4$ meV with an additional 
Gaussian [cf. Eq.~(\ref{eq:L+G})] for data below 8 K. 
The inset shows the magnetic field dependence of
the intensity at $E = 0.4$ meV. 
(b) Constant-$E$ scans along the line 
$\bm{Q} = (\frac{1}{2} \frac{1}{2} L)$ 
taken with $E = 0.75$ meV. 
Solid line is a guide to the eye and dashed line 
is the orientation factor of an anisotropic spin fluctuation 
in the $ab$ plane,  
for data at 1.6 K.
}
\end{figure}
To investigate the energy response around $\bm{k}_2$, 
we performed constant-$Q$ scans at 
$\bm{Q} = (\frac{1}{2} \frac{1}{2} 1)$ 
and time-of-flight measurements with a locus 
approximately along the line 
$(\frac{1}{2} \frac{1}{2} L)$ for $0.8<L<2.2$. 
We note that $\bm{k}_2$ is close to the antiferromagnetic wave vectors 
of Ce(Ni$_{1-x}$M$_{x}$)$_2$Ge$_2$ with M = Pd and Rh
\cite{FakSSC00-Fukuhara02}. 
Figure~\ref{fig:EQscan}(a) shows energy spectra 
in the temperature range $0.1<T<20$ K. 
One can see a pronounced enhancement of the low-energy spin fluctuations
at low temperatures where NFL behavior in bulk 
properties become evident. 
In order to show the relevance of these low-energy spin fluctuations 
to the NFL behavior, we measured the magnetic-field dependence 
of the intensity at $E = 0.4$ meV. 
The inset of Fig.~\ref{fig:EQscan}(a) shows a significant reduction in the 
intensity,  in agreement with the 
recovery of the FL behavior  with applied field
\cite{Steglich96-Julian96}. 
We conclude that the observed enhancement 
of the low-energy spectral weight is at the origin of the 
NFL behavior. 

The energy spectrum at $T = 20$ K is well described 
by the Lorentzian form of Eq.~(\ref{eq:L}) 
with $\Gamma_{\bm{Q}} = 4$ meV, 
as reported in Refs.~\cite{Fak-JPCM,Knopp88}. 
To fit the additional peak structure below 1.5 meV at 
$T<8$ K, we parametrize the data by adding an \textit{ad hoc} Gaussian 
\begin{equation}
\label{eq:L+G}
\text{Im} \chi_{\text{G}} (\bm{Q}, E) =
\delta \chi (\bm{Q}) 
\frac{\sqrt{\pi} E }{ \gamma_{\bm{Q}} } 
\text{e}^{ - (E/\gamma_{\bm{Q}})^2 }
\end{equation}
to Eq.~(\ref{eq:L}). The resulting fits of Eqs.~(\ref{eq:L}) and (\ref{eq:L+G}) give
an excellent description of the data [solid lines in Fig.~\ref{fig:EQscan}(a)].
It is also possible to describe the data by two Lorentzians for 
$E < 1.5$ meV, but the long tail of the second Lorentzian disagrees with 
the data at higher energies. 
Below $T < 1.6$ K, the  energy width 
[half width at half maximum (HWHM)] of the low-energy Gaussian 
(Lorentzian) term is 0.7 (0.45) meV. 
We note that the necessity to include Eq.~(\ref{eq:L+G}) expresses  
the failure of the spin-fluctuation theory \cite{Moriya-Taki} in another way. 

Figure~\ref{fig:EQscan}(b) shows that antiferromagnetic correlations, peaked 
at integer values of $L$ along $\bm{Q} = (\frac{1}{2} \frac{1}{2} L)$ at 
$E = 0.75$ meV, develop only at low temperatures. 
The slow $\bm{Q}$ variation of the intensity can be described by 
an orientation factor $(1 + \hat{Q}_{c}^{2})$ 
[see dashed curve in Fig.~\ref{fig:EQscan}(b)], 
which implies that the spins fluctuate predominantly in the $ab$ plane.
This spin anisotropy is consistent with the 
antiferromagnetic structures of the Pd-doped compounds 
\cite{FakSSC00-Fukuhara02}. 
However it disagrees with the susceptibility measurements 
\cite{Fuku-MetaM}, which 
indicate an Ising-like anisotropy along the $c$ axis 
at $T \sim 50$ K.

\begin{figure}
\begin{center}
\includegraphics[width=6.0cm,clip]{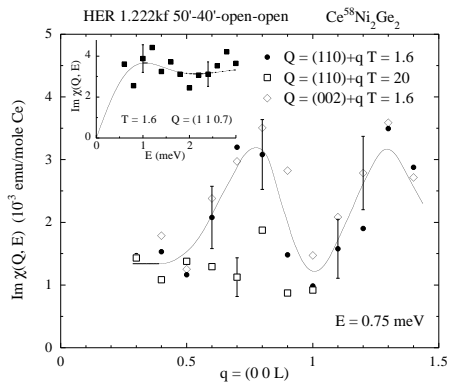}
\end{center}
\caption{
\label{fig:Qscan(00L)}
Constant-$E$ scans along lines 
$\bm{Q} = (11L)$ and $(00L)$ 
taken with $E = 0.75$ meV at $T = 1.6$ and 20 K. 
The solid line is a guide to the eye. 
The inset shows a constant-$Q$ scan at the peak position 
$\bm{Q} = (1,1,0.7)$ and a fit to Eqs.~(\ref{eq:L}) and (\ref{eq:L+G}).
}
\end{figure}
To characterize the spin fluctuation at $\bm{k}_3$, constant-$E$ 
scans along $(11L)$ and $(00L)$ are shown in 
Fig.~\ref{fig:Qscan(00L)}. 
The similar intensities between the $(11L)$ and $(00L)$ 
scans indicate that the spin fluctuations are isotropic. 
 The energy spectrum at the peak position $\bm{Q} = (1,1,0.7)$
(see the inset of Fig.~\ref{fig:Qscan(00L)})
was also parametrized using Eqs.~(\ref{eq:L}) and (\ref{eq:L+G})
assuming $\Gamma_{\bm{Q}} = 4$ meV for the Lorentzian. 
The energy width of the Gaussian is 0.9 meV (HWHM) at 1.6 K. 
This is slightly larger than that of $\bm{k}_2$, 
suggesting that  the spin fluctuations at 
$\bm{k}_3$ have a smaller importance for the NFL behavior. 

\begin{figure}
\begin{center}
\includegraphics[width=7.5cm,clip]{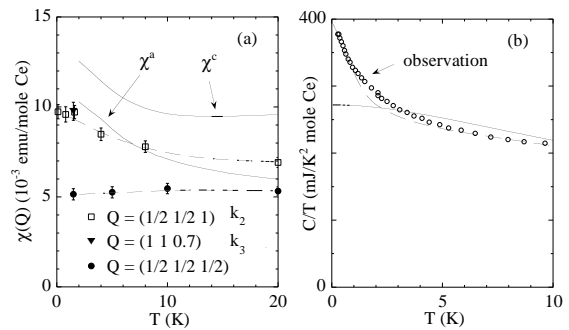}
\end{center}
\caption{
\label{fig:Chi-C/T}
(a) Temperature dependence of the wave-vector dependent susceptibility 
$\chi(\bm{Q})$ and 
uniform susceptibilities $\chi^c$ and $\chi^a$ 
(from Ref.~\cite{Aoki}). 
Error bars of $\chi(\bm{Q})$ include only statistical errors; 
the systematic uncertainty in the absolute normalization
can be up to a factor of 1.5.
Dashed lines are guides to the eye. 
(b) Observed specific-heat coefficient $C/T$ 
(from Ref.~\cite{Koerner00-Cichorek03}) compared with 
model calculations using the SCR spin-fluctuation 
theory \cite{Moriya-Taki}. 
The solid and dashed lines are the SCR evaluations without 
and with the low-energy spin fluctuations, respectively. 
}
\end{figure}
Finally, we compare the present neutron 
data with other measurements. 
The wave-vector dependent susceptibilities 
$\chi(\bm{Q})$ at 
$\bm{Q} = (\frac{1}{2} \frac{1}{2} 1)$ ($\bm{k}_2$), 
$(1,1,0.7)$ ($\bm{k}_3$), 
and 
$(\frac{1}{2} \frac{1}{2} \frac{1}{2})$ \cite{kado03} 
were calculated using the Kramers-Kronig relation from 
$\text{Im} \chi(\bm{Q}, E)$, and 
are shown in Fig.~\ref{fig:Chi-C/T}(a) 
together with the uniform susceptibilities $\chi^c$ and $\chi^a$. 
While the susceptibility $\chi(\bm{Q})$ 
at $\bm{Q} = (\frac{1}{2} \frac{1}{2} \frac{1}{2})$ shows 
a $T$ independent FL behavior, 
$\chi(\bm{k}_2)$ reproduces the upturn at low temperatures 
of $\chi^c$ and $\chi^a$. 

Since spin fluctuations dominate the specific heat 
at low temperatures, one would like to 
ask to what extent the observed low-energy spin fluctuations 
account for the NFL behavior of $C/T$. 
This can be answered semiquantitatively 
by using the self-consistent renormalization (SCR) theory 
of spin fluctuations \cite{Moriya-Taki}, which 
were applied to several heavy fermions \cite{Kambe97}. 
Since $C$ is theoretically calculated from 
$\text{Im} \chi(\bm{Q},E)$ approximated by the Lorentzian form, 
a contribution from the 4 meV spin fluctuation
can be calculated using the SCR technique. 
Figure~\ref{fig:Chi-C/T}(b) shows this part of $C/T$, evaluated assuming 
a $\bm{Q}$ independent $\Gamma_{\bm{Q}} = 4$ meV, by a solid line 
together with the observed $C/T$ \cite{Koerner00-Cichorek03}. 
These show reasonable agreement above $T > 5$ K. 
An estimate of $C/T$ including the low-energy spin fluctuations 
was obtained by replacing the Lorentzian spectral weight with 
the observed data \cite{SCR-excs}. 
It is plotted by a dashed curve in Fig.~\ref{fig:Chi-C/T}(b), 
showing an NFL upturn below $T < 5$ K with almost the same 
magnitude as the observed $C/T$. 
We conclude that the NFL behaviors observed in bulk properties 
are crossover effects due to the 
antiferromagnetic low-energy spin fluctuations. 
From the nondivergent behavior of $\chi(\bm{k}_2)$ 
[see Fig.~\ref{fig:Chi-C/T}(a)] and 
$1/\gamma_{\bm{k}_2}$ [see Fig.~\ref{fig:EQscan}(a)] 
in the limit $T \rightarrow 0$, 
we also conclude that 
the location of CeNi$_2$Ge$_2$ is slightly off the QCP. 
This is supported by the fact that the $E/T$ scaling is not observed in 
CeNi$_2$Ge$_2$; 
most easily seen from the fact that the peak position of the 
low-energy response is independent of $T$ [see Fig.~\ref{fig:EQscan}(a)]. 
In agreement with this interpretation, the recovery of the FL behavior, i.e., 
$C/T =$ const has been reported \cite{Koerner00-Cichorek03} for 
some stoichiometric samples at the lowest temperatures $T< 0.3$ K. 

An aspect of the antiferromagnetic low-energy spin fluctuations 
that cannot be explained by the spin-fluctuation theory \cite{Moriya-Taki} 
was addressed in the itinerant-localized duality theory 
\cite{Duality90-91}. 
The dynamical susceptibility $\chi(\bm{Q},E)$ was derived 
in the theory as 
$\chi(\bm{Q},E)^{-1} = \chi_0(E)^{-1} - \Pi(\bm{Q},E) - J(\bm{Q})$, 
where $\chi_0(E)$ is a local spin susceptibility 
and $J(\bm{Q})$  the Fourier transform of the RKKY interactions. 
The function $\Pi(\bm{Q},E)$ reflects properties of the quasiparticle 
bands, and is usually absorbed into $J(\bm{Q})$ 
by neglecting its $E$ dependence. 
The resulting $\chi(\bm{Q},E)^{-1} = \chi_0(E)^{-1} - J(\bm{Q})$ 
was used as the starting assumption in the spin-fluctuation 
theory \cite{Moriya-Taki}. 
However, the development of a particular quasiparticle band can bring about 
a non-negligible $E$ dependence of $\Pi(\bm{Q},E)$, which was discussed in 
Ref.~\cite{Duality90-91} in connection with 
two kinds of spin fluctuations with energy scales of 5 meV and 0.2 meV 
in the heavy-fermion superconductor UPt$_3$ \cite{Aeppli88}. 
We may speculate that the enhanced low-energy spin fluctuations 
and the deviation from the standard spin-fluctuation description 
of CeNi$_2$Ge$_2$ can be explained in this fashion. 
At present, however, other theoretical scenarios will have to be pursued. 

It is interesting to compare the present results with two other 
compounds that are close to QCP and which have 
been studied in detail by single-crystal neutron scattering: 
CeCu$_{6-x}$Au$_{x}$ with $x_{\text{c}}=0.1$ \cite{Schroder00} and 
Ce$_{1-x}$La$_{x}$Ru$_2$Si$_2$ with $x_{\text{c}}=0.075$ \cite{Raymond97-01}. 
The dynamical susceptibilities of these systems can be, 
at least approximately, described by 
a single Lorentzian [cf. Eq.~(\ref{eq:L})] with 
$\Gamma_{\bm{Q}} \rightarrow 0$ 
at the antiferromagnetic $\bm{Q}$, in agreement with 
the spin-fluctuation theories 
\cite{Moriya-Taki,Hertz76-Millis93}. 
The essential problem of QCP is to determine the singularity, 
which may be different from the mean-field-type solutions 
of the spin-fluctuation theories. 
For CeCu$_{6-x_{\text{c}}}$Au$_{x_{\text{c}}}$ \cite{Schroder00}, 
a detailed study of the divergence revealed a significant deviation 
from the single Lorentzian, 
which led them to propose an extended functional form 
with a non-standard exponent of $\alpha \sim 0.75$. 
On the other hand, for 
Ce$_{1-x_{\text{c}}}$La$_{x_{\text{c}}}$Ru$_2$Si$_2$ \cite{Raymond97-01} 
$\Gamma_{\bm{Q}}$ stays finite in the limit $T \rightarrow 0$. 
In this context, an exactly tuned system, e.g., 
CeNi$_{2-x_{\text{c}}}$Pd$_{x_{\text{c}}}$Ge$_2$ 
\cite{Fukuhara98,Knebel99} with $x_{\text{c}} \sim 0.09$, 
is a promising candidate for studying 
divergent behavior of the Gaussian term of Eq.~(\ref{eq:L+G}). 

In conclusion, we have identified the low-energy spin fluctuations 
that lead to the NFL behavior in CeNi$_2$Ge$_2$. 
They are antiferromagnetic correlations around 
wave vectors $(\frac{1}{2} \frac{1}{2} 0)$ and $(0 0 \frac{3}{4})$ 
with a characteristic energy scale of 0.6 meV. 

We wish to acknowledge T. Moriya for valuable discussions, 
especially on the SCR theory.

\bibliography{ceni2ge2}

\begin{thebibliography}{28}
\bibitem[{SCR()}]{SCR-excs}
\bibinfo{note}{This should be regarded as an order estimation, because the
  replacement has no strict theoretical justification.}
\end{thebibliography}

\end{document}